\pdfoutput=1
\documentclass[reprint,amsmath,amssymb,aip,jcp,groupedaddress]{revtex4-1}

\usepackage[english]{babel}
\usepackage[utf8x]{inputenc}
\usepackage[T1]{fontenc}

\usepackage{graphicx}
\usepackage{ushort}
\usepackage{bm} 
\usepackage[hyperindex,breaklinks]{hyperref}
\usepackage[dvipsnames]{xcolor}

\usepackage[left=2.1cm,right=2.1cm,top=2cm,bottom=1.5cm]{geometry}

\newcommand{\e}{\mathrm{e}}
\renewcommand{\i}{\mathrm{i}}
\newcommand{\xx}{\mathbf{x}}
\newcommand{\rr}{\mathbf{r}}
\newcommand{\jj}{\mathbf{j}}
\newcommand{\jpara}{\jj_\mathrm{para}}
\newcommand{\jdia}{\jj_\mathrm{dia}}
\newcommand{\EE}{\mathbf{E}}
\newcommand{\FF}{\mathbf{F}}
\newcommand{\kk}{\mathbf{k}}

\renewcommand{\AA}{\mathbf{A}}
\newcommand{\Ax}{\mathbf{A}_\mathrm{x}}

\newcommand{\vxc}{v_\mathrm{xc}}
\newcommand{\vx}{v_\mathrm{x}}
\newcommand{\vH}{v_\mathrm{H}}
\newcommand{\vHx}{v_\mathrm{Hx}}
\newcommand{\vHxc}{v_\mathrm{Hxc}}
\newcommand{\uHxc}{u_\mathrm{Hxc}}
\newcommand{\vc}{v_\mathrm{c}}
\newcommand{\vxalpha}{v_\mathrm{x}^{\mathrm{X}\alpha}}
\newcommand{\Axc}{\mathbf{A}_\mathrm{xc}}
\newcommand{\FH}{\FF_\mathrm{H}}
\newcommand{\Fx}{\FF_\mathrm{x}}
\newcommand{\FHxc}{\FF_\mathrm{Hxc}}

\newcommand{\nablaleft}{\overset{\leftarrow}{\nabla}}
\renewcommand{\d}{\,\mathrm{d}} 
\DeclareRobustCommand{\onehalf}{\textstyle \frac{1}{2}}
\DeclareMathOperator{\sign}{sign}

\begin{document}

\title{Force Balance Approach for Advanced Approximations in Density Functional Theories}

\author{Mary-Leena M. Tchenkoue}
\email{mary-leena.tchenkoue@mpsd.mpg.de}
\affiliation{Max Planck Institute for the Structure and Dynamics of Matter, Luruper Chaussee 149, 22761 Hamburg, Germany}
\author{Markus Penz}
\email{markus.penz@mpsd.mpg.de}
\affiliation{Max Planck Institute for the Structure and Dynamics of Matter, Luruper Chaussee 149, 22761 Hamburg, Germany}
\author{Iris Theophilou}
\affiliation{Max Planck Institute for the Structure and Dynamics of Matter, Luruper Chaussee 149, 22761 Hamburg, Germany}
\author{Michael Ruggenthaler}
\affiliation{Max Planck Institute for the Structure and Dynamics of Matter, Luruper Chaussee 149, 22761 Hamburg, Germany}
\author{Angel Rubio}
\affiliation{Max Planck Institute for the Structure and Dynamics of Matter, Luruper Chaussee 149, 22761 Hamburg, Germany}
\affiliation{Center for Computational Quantum Physics, The Flatiron Institute, New York, NY 10010, USA}

\date{\today}

\begin{abstract}
We propose a systematic and constructive way to determine the exchange-correlation potentials of density-functional theories including vector potentials. The approach does not rely on energy or action functionals. Instead it is based on equations of motion of current quantities (force balance equations) and is feasible both in the ground-state and the time-dependent setting. This avoids, besides differentiability and causality issues, the optimized-effective-potential procedure of orbital-dependent functionals. We provide straightforward exchange-type approximations for different density functional theories that for a homogeneous system and no external vector potential reduce to the exchange-only local-density and Slater X$\alpha$ approximations.
\end{abstract}

\maketitle

\section{Introduction}
It is hard to doubt that the Achilles heel of density-functional theory (DFT) is the lack of highly accurate and at the same time generally-applicable approximate functionals. Following the famous Jacob's ladder of Perdew~\cite{Perdew_Jacob}, the most advanced approximations available to date for Kohn--Sham DFT are orbital-dependent functionals. However, they have the major drawback that they make a numerically costly optimized-effective-potential (OEP) procedure~\cite{OEP_1} necessary to determine the exchange-correlation (xc) potential. In special cases, such as exact-exchange-type approximations, this drawback can be mitigated somewhat if one switches from Kohn--Sham DFT to generalized Kohn--Sham DFT, where a non-local Hartree--Fock exchange potential is allowed besides the local xc potential \cite{GKS, Orbital-dependent-functionals}. Furthermore, there exist attempts to directly construct approximate xc potentials without the OEP procedure, by ensuring instead that the resulting xc potentials indeed correspond to functional derivatives~\cite{Staroverov_func_der, Staroverov_pot_driven}. However, there are several further subtle problems associated with using energy or action (in time-dependent density-functional theory (TDDFT)) functionals in order to construct approximate xc potentials. In principle, the exact ground-state energy expressions are non-differentiable with respect to the density~\cite{Lammert2007} and regularization procedures need to be employed~\cite{Kvaal2014,KSpaper2018}. In the time-dependent case, the action-functional approach suffers from causality problems~\cite{VANLEEUWEN2001}, and makes necessary the use of either the Keldysh time contour~\cite{vanLeeuwen1998}, the equivalent use of time-ordered superoperators~\cite{Mukamel2005}, or extra temporal boundary conditions~\cite{Vignale2008}. 

If we move beyond the one-particle density as a basic variable and, for instance, consider also the current density, even more issues arise. Besides the fact that there are relatively few current-density functionals available (see e.g.\ Ref.~\onlinecite{CDFT_functional_Teale} and references therein), there is an ambiguity concerning which current-density quantity one should use \cite{Tellgren2012,Laestadius2014}. Indeed, if the physical current density is used in the ground-state case, the usual variational approach is not even applicable and a definition of the xc potential in the usual way is not available any more \cite{laestadius2015nonexistence}. To overcome this restriction one commonly employs the paramagnetic current density instead to set up a current-density-functional theory (CDFT) for ground states, which makes the theory gauge dependent~\cite{vignale-rasolt-geldart}. In the time-dependent case, however, the gauge-independent physical current density is usually used to formulate a time-dependent current-density-functional theory (TDCDFT). Thus approximations from CDFT are in general not trivially connected to TDCDFT.

In practice most of these subtle issues are disregarded, and instead of using purpose-built approximate functionals for different cases, the abundant ground-state density-functional approximations are applied to time-dependent situations (adiabatic approximation) and to cases including magnetic fields, where the current density becomes a necessary basic functional variable, alike. 

Besides the usual energy or action-functional approaches, the foundations of TDDFT and TDCDFT already suggest that there is a direct way towards the xc potentials based on the equations of motion (EOM) of the physical current density (force balance equation) and the EOM of the one-particle density (continuity equation). This ``hydrodynamical'' perspective was already proposed in the pioneering work of Runge and Gross~\cite{runge1984density} and it was discussed thoroughly by Tokatly~\cite{tokatly2005quantum,tokatly2005quantum2,tokatly2009time,Tokatly2012} and Vignale~\cite{TDCDFT_Vignale_proof,vignale1997time,Vignale2012}. Based on this idea, several approximations to the time-dependent xc potentials have been proposed~\cite{vignale1997time,tokatly2005quantum2,tao2006time,ruggenthaler2009local}. More recent works include approximations for multi-determinantal Kohn--Sham TDDFT~\cite{fuks2016time} and for standard (single-determinantal) Kohn--Sham TDDFT~\cite{lacombe2019density}. However in the context of ground-state problems, besides Refs.~\onlinecite{ruggenthaler2009local, keDFT_lattice}, EOM-based xc potentials have not been investigated in more detail as an alternative to energy-functional-based approximations.

In this work we introduce a systematic way to employ problem-adopted EOMs for obtaining exact determining relations for xc potentials that link the interacting and non-interacting systems of the Kohn--Sham construction~\cite{Kohn-Sham} for different density-functional theories. In some settings these relations were already discussed~\cite{tokatly2005quantum2,Tokatly2012} and their integrated forms are known as the zero-force and zero-torque constraint for $\vxc$, which exist both for the static~\cite{Perdew_scaling} and time-dependent scalar xc potential~\cite{Vignale_zero_force}. We focus here on the more general Hamiltonian of TDCDFT~\cite{Ghosh-Dhara, TDCDFT_Vignale_proof} and the resulting EOM of the current density. The determining relations that follow provide a unifying framework for the above mentioned different density-functional theories. For general time-dependent external scalar and vector potentials we arrive at exact equations for $\vxc$ and $\Axc$, and setting the vector potential to zero we find the corresponding equation for the time-dependent $\vxc$ of TDDFT. On the other hand, for static external scalar and vector potentials and considering stationary states, we are in the setting of CDFT, where we get exact equations for $\vxc$ and $\Axc$ irrespective of whether we choose the paramagnetic or the physical current density as basic variables (yet the respective equations will be different). Setting now also the vector potential to zero and making the potential time-independent, we obtain a pointwise relation determining the $\vxc$ of ground-state DFT (as in the original Hohenberg--Kohn formulation). These exact relations allow us to straightforwardly establish problem-adopted exact-exchange-type approximations without the above mentioned drawbacks of energy- and action-functional-based approaches. At the same time they clarify the interrelations of the xc potentials of different density-functional theories. Finally we highlight the connection of such EOM-based approximations to the usual energy-functional-based approximations by showing how in the case of a uniform electron gas the exact-exchange-type approximations reduce to the well-known exchange-only local-density approximation (LDA). This connection also provides a new perspective on the Slater exchange potential $\vxalpha[n]$~\cite{slater1951}.

The paper is organized as follows: In Sec.~\ref{sec:level2} we discuss the EOM of the first-order reduced density matrix which provides all pointwise determining relations for one-body quantities. We then focus on the resulting EOM of the current density (physical and paramagnetic), the ``force balance equation'', to obtain determining relations for the xc potentials in Sec.~\ref{sec:exx} and subsequent functional approximations of exact-exchange type. In Sec.~\ref{sec:LDA} we show how EOM-based approximations in the homogeneous case reduce to the usual exchange LDA potential. In Sec.~\ref{sec:conclusions} we summarize our results and discuss implications.

\section{\label{sec:level2}Equations of Motion}

In this section we present the fundamental EOM of the first-order reduced density matrix from which all other one-body EOMs can be deduced. We then identify the two simplest EOMs, the one of the one-particle density and the one of the current density, that allow to establish a direct relation between the basic functional variables (one-particle density, paramagnetic or physical current density) and the corresponding potentials (scalar and/or vector potential).

Let us first set the stage and introduce the many-electron Hamiltonian we consider in Hartree atomic units ($e = \hbar = m_e = (4\pi\epsilon_0)^{-1} = 1$),
\begin{align}
H = \underbrace{\sum_{k=1}^N h(\rr_k,t)}_{T+V}  + \underbrace{\sum_{k>l} w(\rr_k-\rr_l)}_W,
\label{hamiltonian}
\end{align}
where $w(\rr_k-\rr_l)$ is the electron-electron interaction term and the single-particle Hamiltonian reads
\begin{align}
h(\rr,t)&=\tfrac{1}{2}(-\i\nabla-\AA(\rr,t))^2 +v(\rr,t). \label{ham-magnetic-single-particle}
\end{align}
We consider the magnetic Schrödinger equation with external scalar potential $v(\rr,t)$ and external vector potential $\AA(\rr,t)$ to be able to cover the setting of CDFT where $\AA$ is the conjugate variable for the current density.
Next, we see how to reduce complexity while still retaining the full and exact information about at least some properties of the system. Given that only one- and two-body operators are present in Eq.~\eqref{hamiltonian}, a reduced form of the wave function that still gives exact results for the expectation values of the Hamiltonian would have to include at least two different particle coordinates. Such a reduced form is the $p$\textsuperscript{th}-order reduced density matrix ($p$-RDM), $p\in(1,\ldots,N),$
\begin{equation}
\begin{aligned}
\rho_{(p)}&(\rr_1, \ldots, \rr_p,\rr_1', \ldots, \rr_p',t)\\
=& \frac{N!}{p!(N-p)!} \sum_{\substack{s_1 \ldots s_p \\ s'_1 \ldots s'_p}}\int \Psi(\xx_1,\ldots,\xx_p,\xx_{p+1},\ldots,\xx_N,t)\\
& \Psi^{*}(\xx'_1,\ldots,\xx_p',\xx_{p+1},\ldots,\xx_N,t) \d\xx_{p+1}\ldots\d\xx_N,
\label{p-RDM}
\end{aligned}
\end{equation}
where the coordinate $\xx_i = (\rr_i, s_i)$ comprises both space and spin coordinates. Note that consequently an integral over a combined coordinate $\xx_i$ includes both, a $\mathbb{R}^3$ integral over the space variable $\rr_i$ and a sum over the spin degrees of freedom $s_i \in \{\uparrow,\downarrow\}$.
The $\Psi(\xx_1,\ldots,\xx_p,\xx_{p+1},\ldots,\xx_N,t)$ is a general many-electron wave function, which could, for instance, correspond to the ground state of the Hamiltonian from Eq.~\eqref{hamiltonian} (there will be no dependence of time in this case). Further, we denote the diagonal of the 1-RDM as 
\begin{equation}
\label{eq-density}
n(\rr_1,t) = \rho_{(1)}(\rr_1,\rr_1,t) = N \sum_{s_1}\int |\Psi|^2 \d\xx_2\ldots\d\xx_N, 
\end{equation}
which gives the one-particle density, the functional variable that substitutes the wave function in DFT.

\subsection{EOM for the reduced density matrix (1-RDM)}

Since the expectation value of any one-body operator can be calculated solely from the 1-RDM~\cite[Sec.~1.7]{stefanucci2013nonequilibrium}, we first give the EOM for the 1-RDM as the general case,
\begin{equation}\label{EOM-1RDM}
\begin{aligned}
&\i\partial_t \rho_{(1)}(\rr_1,\rr'_1,t)=\left[h(\rr_1,t)-\overline{h}(\rr'_1,t)\right]\rho_{(1)}(\rr_1,\rr'_1,t)\\
&+2\int \left[w(\rr_2-\rr_1)-w(\rr_2-\rr'_1)\right] \rho_{(2)}(\rr_1, \rr_2, \rr'_1,\rr_2,t) \d \rr_2.
\end{aligned}
\end{equation}
The overline indicates that the operator needs to be complex conjugated.
The EOM for the 1-RDM can also be found in Ref.~\onlinecite{Giesbertz2012}, however they use a different normalization convention for the 2-RDM.
From Eq.~\eqref{EOM-1RDM} we can derive the EOMs of any one-body observable. EOMs for many-body observables on the other hand are not useful for us, as the usual Kohn--Sham Hamiltonian by construction does not involve many-particle operators.

As an example, we derive the EOM for the one-particle density (the continuity equation) from the EOM of the 1-RDM above,
\begin{equation}
\i\partial_t n(\rr_1,t)= \left.\i\partial_t \rho_{(1)}(\rr_1,\rr_1',t)\right|_{\rr'_{1}=\rr_{1}}.
\end{equation}
Since one sets $\rr'_1 = \rr_1$ after acting with the operators in the right hand side of Eq.~\eqref{EOM-1RDM} the whole interaction part drops out as well as the terms including the external potential $v$, just like mentioned before. 
The remaining parts give
\begin{equation}
\begin{aligned}
\i\partial_t n(\rr_1,t)
=\;& \frac{1}{2}\left[(\i\nabla+\AA(\rr_1,t))^2 \right. \\
&\left.\left.-(\i\nabla'-\AA(\rr'_1,t))^2\right]\rho_{(1)}(\rr_1,\rr'_1,t)\right|_{\rr'_1=\rr_1},
\end{aligned}
\label{continuity-pre}
\end{equation}
where $\nabla$ acts on the coordinate $\rr_1$ and $\nabla'$ on $\rr'_1$.
As a result we get the continuity equation
\begin{equation}
\partial_t n(\rr_1,t)= -\nabla \cdot \jj(\rr_1,t),
\label{continuity}
\end{equation}
where we have introduced the (physical) current density that can be split into a paramagnetic and a diamagnetic contribution,
\begin{equation}
\begin{aligned}
&\jj(\rr_1,t) \\
&= \left.\frac{1}{2\i}\left[(\nabla-\i\AA(\rr_1,t))-(\nabla'+\i\AA(\rr'_1,t))\right]\rho_{(1)}(\rr_1,\rr'_1,t)\right|_{\rr'_1=\rr_1}\\
&=N \Im \sum_{s_1} \int \Psi^* \nabla \Psi \d \xx_2\ldots\d\xx_N - \AA(\rr_1,t) n(\rr_1,t)\\
&= \jpara(\rr_1,t) + \jdia(\rr_1,t).
\label{current_def}
\end{aligned}
\end{equation}

As we will discuss later, the continuity equation provides an interesting identity for the xc vector potential in Eq.~\eqref{eq:remarkable}. However, this does not provide a useful connection between the interacting and the non-interacting system that could be used for functional construction, since no relation between the xc vector potential and the current quantity shows up. Still this identity can serve as a further condition for functional approximations. In the case of an EOM for the current density (force balance equation), the right hand side will also involve the (different) potentials, and it can thus provide us with useful relations for functional construction in both, the ground-state and the time-dependent setting.

\subsection{EOM for the current density (force balance equation)}
\label{sec:eom-current}

In order to determine the EOM for the current density we take the time derivative of Eq.~\eqref{current_def} and use the EOM of the 1-RDM \eqref{EOM-1RDM}. Setting $\rr_1'=\rr_1$ leads to the cancellation of a few terms and making use of the continuity equation \eqref{continuity} yields
\begin{equation}
\begin{aligned}
\partial_t \jj(\rr_1,t) =& - n(\rr_1,t) \partial_t \AA(\rr_1,t) + \frac{1}{2\i}\left[(\nabla-\i\AA(\rr_1,t))\right. \\
&\left.\left. - (\nabla'+\i\AA(\rr'_1,t))\right]\partial_t\rho_{(1)}(\rr_1,\rr'_1,t)\right|_{\rr'_1=\rr_1} \\
=& -\frac{1}{4} (\nabla - \nabla') \left[(\i\nabla+\AA(\rr_1,t))^2\right. \\
&-\left.\left.(\i\nabla'-\AA(\rr'_1,t))^2\right]\rho_{(1)}(\rr_1,\rr'_1,t)\right|_{\rr'_1=\rr_1} \\
&+ \AA(\rr_1,t) (\nabla \cdot \jj(\rr_1,t)) \\
&- (\nabla v(\rr_1,t) + \partial_t \AA(\rr_1,t))n(\rr_1,t)  \\
&- 2\int (\nabla w(\rr_2-\rr_1)) \rho_{(2)}(\rr_1, \rr_2, \rr_1,\rr_2,t)\d \rr_2.
\end{aligned}
\end{equation}
At this point it is already possible to identify different force terms, like the interaction force,
\begin{equation}\label{eq-interaction-force}
\FF_W[\Psi](\rr_1,t)=- 2\!\int (\nabla w(\rr_2-\rr_1)) \rho_{(2)}(\rr_1,\rr_2,\rr_1,\rr_2,t) \d \rr_2,
\end{equation}
and the (gauge-independent) contribution from the electric field $\EE = - (\nabla v + \partial_t \AA)$ that couples exclusively to the density. If we separate the parts including the vector potential $\AA$ in the first term involving the 1-RDM, the remaining terms give rise to the kinetic force term,
\begin{equation}\label{eq-kinetic-force}
\begin{aligned}
&\FF_T[\Psi](\rr_1,t) = \left.\frac{1}{4} (\nabla - \nabla')(\nabla^2 - \nabla'^2)\rho_{(1)}(\rr_1,\rr'_1,t)\right|_{\rr'_1=\rr_1}  \\
&=\frac{N}{2}\Re \sum_{s_1}\int \left( \left(-\nabla^2\Psi^*\right) \nabla \Psi + \Psi^* \nabla \nabla^2 \Psi \right) \d\xx_2\ldots\d\xx_N.
\end{aligned}
\end{equation}
All the remaining parts can be combined in order to yield the contribution from the Lorentz force as well as further internal forces involving the vector potential,
\begin{equation}
\begin{aligned}
\partial_t \jj =& -n\overbrace{(\nabla v + \partial_t \AA)}^{-\mathbf{E}} + \jj \times \overbrace{(\nabla \times \AA)}^{\mathbf{B}} + (\AA \otimes \jj)\nablaleft \\
&+ ((\jj + n\AA) \otimes \AA)\nablaleft + \FF_T[\Psi] + \FF_W[\Psi].
\end{aligned}
\label{eq-force-balance}
\end{equation}
This is the EOM for the current density, the \emph{force balance equation}, that can also be found in index notation in Ref.~\onlinecite[Eq.~(3.37)]{stefanucci2013nonequilibrium}. In order to be able to give the compressed form above we use the dyadic (or outer) product defined as
\begin{equation}
\left((\mathbf{a} \otimes \mathbf{b}) \nablaleft\right)_i = \sum_j \frac{\partial}{\partial r_j} \mathbf{a}_i\textbf{b}_j,
\end{equation}
as well as the vector identities
\begin{align}
(\AA \otimes \jj)\overset{\leftarrow}{\nabla} &= (\nabla \cdot \jj) \AA + (\jj \cdot  \nabla)\AA, \label{eq:vec-id-1} \\
\jj \times (\nabla \times \AA) &= (\nabla \otimes \AA)\jj - (\jj \cdot \nabla)\AA. \label{eq:vec-id-2}
\end{align}

One can immediately see that the two first terms on the right hand side of Eq.~\eqref{eq-force-balance} in their integrated form give the Lorentz force, Ref.~\onlinecite[Eq.~(3.38)]{stefanucci2013nonequilibrium}. Further, $\int \FF_T \d \rr =\int \FF_W \d\rr=0$ holds, as these terms can be rewritten into divergence form by a non-trivial manipulation illustrated in Ref.~\onlinecite{tokatly2005quantum}. Then Gauss' theorem and the fact that the corresponding surface integrals will vanish at infinity tell us that the integrals are zero.
 In the case of $\FF_W$ it can alternatively be argued that the full space integral equals zero due to the symmetry of the interaction potential $w(\rr_2-\rr_1)$ upon interchange of $\rr_1$ and $\rr_2$.
 For the remaining two terms the divergence form is obvious. This means integration over the whole space gives the classical force expression with all quantum mechanical contributions canceled as purely internal effects.
 
The critical feature of the current density EOM is that it connects the density quantities $(n,\jj)$ with the steering external potentials $(v,\AA)$. This connection was already employed in the classical Runge--Gross proof \cite{runge1984density}, but in a setting without vector potentials. It enables us to look for the correct potentials that yield given densities (density-potential map) or, in the case of the Kohn--Sham construction that we study in Sec.~\ref{sec:exx}, rather for the difference between the necessary potentials for a non-interacting and an interacting system (xc potential).

Calculating the corresponding EOM for the paramagnetic current density gives
\begin{equation}
\begin{aligned}
\partial_t \jpara = &-n \nabla v + (\nabla \otimes \AA)(\jpara - n\AA) \\
&+ (\jpara \otimes \AA)\nablaleft + \FF_T[\Psi] + \FF_W[\Psi].
\end{aligned}
\label{eq-force-balance-para}
\end{equation}

On a first glance the EOMs~\eqref{eq-force-balance} and \eqref{eq-force-balance-para} do not seem to give any useful information when considering ground states, however, as we will show in Sec.~\ref{sec:exx}, this is not the case.

\section{Exact xc potentials from the force balance equation}
\label{sec:exx}

In order to be able to make the usual Kohn--Sham construction~\cite{Kohn-Sham} we assume two systems, one that is described by a Hamiltonian of the form of Eq.~\eqref{hamiltonian} with interactions, and another one with no interaction (Kohn--Sham system) that is described by
\begin{equation}
H_s = \sum_{k=1}^N \left( \frac{1}{2}(-\i\nabla-\AA_s(\rr_k,t))^2 +v_s(\rr_k,t)\right).
\label{KS_Ham}
\end{equation}
The basic idea of the Kohn--Sham construction is now that $(v_s,\AA_s)$ are chosen such that the wave function of the non-interacting system $\Phi$ generates the same fundamental functional variables (one-particle density and/or current density) as the wave function of the interacting system $\Psi$. In the most simple setting, that of standard DFT, where no vector potential is present, one chooses the one-particle densities $n[\Psi] = n[\Phi] = n$ to agree. If a vector potential is added as a further external quantity, some form of the current density (paramagnetic or total) is additionally used to describe the systems. 
Independent of what is chosen on the side of densities to describe the electronic system and what is chosen as an external potential, we will always assume the possibility of a well-defined density-potential mapping for the interacting and the non-interacting system.
Yet it must be noted that delicate $v$-representability issues that are still not fully resolved arise in all cases~\cite{vanleeuwen2003density}. On the other hand, for ground states there is a viable way of how to work with Lieb's convex-conjugate formulation~\cite{Lieb1983} and a special regularization procedure, which \emph{always} guarantees the existence of a well-defined density-potential mapping~\cite{Kvaal2014,KSpaper2018}. This method has been applied specifically to CDFT \cite{MY-CDFTpaper2019} and to show convergence of the Kohn--Sham iteration scheme \cite{KSconvergence2019}.

Next we discuss the derivation of determining expressions for the xc potential, the difference between the potentials belonging to the interacting and non-interacting system, again in different settings: for ground-states and for time-dependent systems including vector potentials. 
%
Note, however, that we need to match EOMs of quantities that can in principle be reproduced by a non-interacting system. For instance, in the case of 1-RDM functional theory~\cite{Gilbert} it is not possible to match the time derivative of an interacting 1-RDM with a non-interacting one, as for zero temperature an interacting 1-RDM cannot even be reproduced by a non-interacting one.

\subsection{(TD)DFT xc potential}
\label{sec:DFT_xc}

%

Let us first discuss the standard case of no external vector potential (thus no magnetic field), which corresponds to the setting of (TD)DFT.
The fact that we can in principle reproduce the density of the interacting system with a non-interacting auxiliary Kohn--Sham system (with a unique $v_s$) is in this case based on the original Hohenberg--Kohn theorem~\cite{Hohneberg-Kohn} that states the existence of a well-defined density-potential map. If both wave functions are ground states of their respective systems (standard DFT setting), the current density EOMs \eqref{eq-force-balance} of both systems are equal to zero. Assuming that the corresponding one-particle densities $n$ are identical we have
\begin{align}
&\partial_t \jj[\Psi] = -n\nabla v + \FF_T[\Psi] + \FF_W[\Psi]=0, \label{interac-force-bal-eq}\\
&\partial_t \jj_s[\Phi] = -n\nabla v_s + \FF_T[\Phi]=0.
\label{noninterac-force-bal-eq}
\end{align}
It is somewhat unusual that we employ an equation for the current density, since with a scalar potential we can only control the density not the current. However, since only the time-derivatives of the current densities appear, which are zero by construction, this does not pose a problem (in TDDFT, where the time-derivative of the current densities are non-zero, it would constitute a problem). Subtracting Eqs.~\eqref{interac-force-bal-eq} and \eqref{noninterac-force-bal-eq} grants access to the Hartree-exchange-correlation potential that is defined as $\vHxc = v_s - v$,
\begin{equation}\label{eq-Hxc-potential}
n\nabla \vHxc = -\FHxc[\Phi,\Psi] = \FF_T[\Phi] - \FF_T[\Psi] - \FF_W[\Psi].
\end{equation}
This equation as a route towards the $\vHxc$ potential was considered before in Ref.~\onlinecite[Eq.~(19)]{tokatly2005quantum2}, and Ref.~\onlinecite[Eqs.~(24) and (25)]{ruggenthaler2009local}, involving the internal stress tensors of both systems.
Since Eq.~\eqref{eq-Hxc-potential} holds \emph{exactly}, it also tells us that $\FHxc[\Phi,\Psi]/n$ is a pure gradient field, having only longitudinal contributions, if $\Phi$ and $\Psi$ are ground-state wave functions to a Schrödinger equation without vector potential. This situation changes in a non-stationary setting that will be considered below. It is customary to expand $\vHxc$ from Eq.~\eqref{eq-Hxc-potential} into a Hartree-exchange and a remaining correlation part,
\begin{align}
    n\nabla \vHxc &= \underbrace{-\FF_W[\Phi]}_{\to \vHx} + \underbrace{\FF_T[\Phi] - \FF_T[\Psi] + \FF_W[\Phi] - \FF_W[\Psi]}_{\to \vc}, \label{eq:vHxc-parts}\\
    n\nabla \vHx &= -\FF_W[\Phi], \label{eq:def-vHx}\\
    n\nabla \vc &= \FF_T[\Phi] - \FF_T[\Psi] + \FF_W[\Phi] - \FF_W[\Psi]. \label{eq:def-vc}
\end{align}
The Hartree-exchange potential $\vHx$ from Eq.~\eqref{eq:def-vHx} is defined by a Slater-determinant wave function of the KS system $\Phi$ alone and can thus be already counted as a full-fledged orbital-dependent approximation to the xc potential, called the \emph{local-exchange} approximation~\cite{ruggenthaler2009local}. That this potential corresponds to the functional derivative of the well-known exchange approximation of the exchange-energy functional~\cite{OEP_1} is not clear apriori. Only for the case of a spin-singlet two-particle system the EOM-based local-exchange approximation has been shown to coincide with the energy-based exchange approximation~\cite{ruggenthaler2009local}. In this work we will further clarify the connection between the local-exchange approximation and energy-based approximations by showing that in the limit of a homogeneous system the local-exchange approximation reduces to the exchange-only LDA in Sec.~\ref{sec:LDA}. The remaining correlation part $\vc$ captures all further differences between the interacting and non-interacting system. For a further distinction between Hartree and exchange parts see also Eq.~\eqref{eq-interaction-force-single-slater} in Sec.~\ref{sec:LDA}. 

Note that we could also derive a defining relation like Eq.~\eqref{eq-Hxc-potential}, that then also works in a time-dependent setting, from considering the second time derivative of the one-particle density instead of the EOM for the current density. Starting from the continuity equation \eqref{continuity} and taking one further time derivative one has $\partial^2_t n=-\nabla \cdot \partial_t \jj$ and thus the current density EOM appears again. Instead of using the current density EOMs set to zero, we employ the corresponding equations for the second time derivative of the densities (that are again required to be identical at all times),
\begin{align}
&\partial_t^2 n = \nabla(n\nabla v) - \nabla( \FF_T[\Psi] + \FF_W[\Psi] ),\\
&\partial_t^2 n = \nabla(n\nabla v_s) - \nabla\FF_T[\Phi].
\end{align}
Subtraction of the two equations above then gives the divergence of Eq.~\eqref{eq-Hxc-potential},
\begin{equation}\label{eq-Hxc-potential-2}
\nabla(n\nabla \vHxc) = -\nabla\FHxc[\Phi,\Psi].
\end{equation}
That such an equation can indeed be efficiently solved for $\vHxc$ was demonstrated numerically in Ref.~\onlinecite{Nielsen2018}.

We thus see that we could alternatively have taken the divergence of the local-force-equation expression \eqref{eq-Hxc-potential} that we used for ground states right away. In the static case Eqs.~\eqref{eq-Hxc-potential} and \eqref{eq-Hxc-potential-2} share exactly the same information about $\vHxc$ and thus also will lead to the same approximations. Yet, Eq.~\eqref{eq-Hxc-potential} cannot be used in the time-dependent setting, since a scalar potential cannot control the full current density (only the longitudinal part) and thus the EOMs of the interacting and the non-interacting system cannot be made to match. In TDDFT we rely on the Runge--Gross theorem~\cite{runge1984density} and its extensions~\cite{vanleeuwen1999,ruggenthaler2011global,ruggenthaler2015topical-review} to control the time-dependent densities and make them match at all times. Consequently Eq.~\eqref{eq-Hxc-potential-2} can be derived also in the time-dependent case and the local-exchange approximation becomes (see also Ref.~\onlinecite[Eq.~(24)]{ruggenthaler2009local})
\begin{align}\label{eq-td-Hx-potential}
\nabla(n\nabla \vHx) = -\nabla\FF_W[\Phi].
\end{align} 
We can, however, no longer infer that $\FHxc/n$ is purely longitudinal. That is why in the alternative derivation based on the EOM for the ground-state current density we needed to employ the divergence to arrive at Eq.~\eqref{eq-Hxc-potential-2}. Its transverse component will show up explicitly as soon as also the current density $\jj$ is to be matched in the CDFT setting, like exercised in Sec.~\ref{Sec:CDFT_xc} and Sec.~\ref{Sec:paraCDFT_xc}.

\subsection{(TD)CDFT xc potentials}
\label{Sec:CDFT_xc}

CDFT is a natural generalization of DFT to cases including magnetic fields. The corresponding complementary density quantity to the external potential $\AA$ that appears in the Hamiltonian of Eq.~\eqref{ham-magnetic-single-particle} is then the current density, but different forms of the current could be chosen to get determining equations for the exact xc potentials.
The usual definition of the xc potentials is as the functional derivatives of the energy expression with respect to the one-particle density and current density, while keeping the potentials fixed~\cite{Vignale-Rasolt-1987,Vignale-Rasolt-1988}. Yet since the vector potential explicitly appears in the expression for the physical current density (see Eq.~\eqref{current_def}), this approach is no longer viable and it is usually argued that instead the paramagnetic current density must be taken as a corresponding density quantity. For the same reason only the paramagnetic current density fits into Lieb's convex-conjugate formulation \cite{Lieb1983} that was adopted to ground-state CDFT in Refs.~\onlinecite{Tellgren2012,MY-CDFTpaper2019}. Exchange-correlation potentials from matching paramagnetic current densities and the closely related vorticity are discussed in Sec.~\ref{Sec:paraCDFT_xc}. But first we consider the physical current density in order to specify the systems and derive the respective determining equations for the xc potentials. This would not be straightforwardly possible with an energy-based approach due to the explicit dependence of the current density on the potentials.

For convenience we repeat Eq.~\eqref{eq-force-balance} for both, the interacting and the non-interacting systems,
\begin{align}
\partial_t \jj =& -n(\nabla v + \partial_t\AA) + \jj \times (\nabla \times \AA) + (\AA \otimes \jj)\nablaleft \nonumber\\
&+ ((\jj + n\AA) \otimes \AA)\nablaleft + \FF_T[\Psi] + \FF_W[\Psi], \label{eq:force-balance-int} \\
\partial_t \jj =& -n(\nabla v_s + \partial_t\AA_s) + \jj \times (\nabla \times \AA_s) + (\AA_s \otimes \jj)\nablaleft \nonumber\\
&+ ((\jj + n\AA_s) \otimes \AA_s)\nablaleft + \FF_T[\Phi]. \label{eq:force-balance-non-int}
\end{align}
That we can indeed match the currents of both systems in the time-dependent case relies on Vignale's extension of the Runge--Gross theorem to TDCDFT~\cite{TDCDFT_Vignale_proof}. For the ground-state case no such theorem that extends the Hohenberg--Kohn result to the physical current density is available to date~\cite{Tellgren2012,Laestadius2014,laestadius2015nonexistence}. In the following we tacitly assume that an appropriate density-potential mapping exists. That there is some hope that such a mapping can be established is also based on recent positive results for specialized cases. These include kinetic-energy DFT on a lattice, where the kinetic-energy density depends explicitly on the spatially-dependent mass~\cite{keDFT_lattice}, DFT based on the Maxwell--Schrödinger equation including an internal magnetic field (MDFT)~\cite{Tellgren2018}, and DFT for coupled matter-photon systems (QEDFT)~\cite{QEDFT}. Apart from the physical current density $\jj$ we will also match the one-particle density $n$ of the interacting and non-interacting system. For the time-dependent setting, due to the availability of the continuity equation, the matching of the densities just follows from the matching of the physical currents, as we discuss in the last paragraph of this subsection. For the ground-state case, the current density and the one-particle density have to be controlled independently, since the time-derivative of the density that appears in the continuity equation will be zero naturally.
Subtracting Eq.~\eqref{eq:force-balance-int} from Eq.~\eqref{eq:force-balance-non-int} gives the determining equation for the xc potentials $\vHxc = v_s-v$ and $\Axc = \AA_s-\AA$ by
\begin{equation}\label{eq-xc-potentials}
\begin{aligned}
n(\partial_t \Axc &+ \nabla \vHxc) = \jj \times (\nabla \times \Axc) + (\Axc \otimes (\jj+n\AA) )\nablaleft  \\
&+((\jj+n\AA+n\Axc) \otimes \Axc)\nablaleft - \FHxc[\Phi,\Psi],
\end{aligned}
\end{equation}
with $\FHxc$ defined as in Eq.~\eqref{eq-Hxc-potential}. Although this equation was considered in Ref.~\onlinecite[Eq.~(21)]{tokatly2005quantum2} before, the dependence of the internal-force terms on the xc vector potential was not made explicit. It is now possible to determine $\vHxc$ and $\Axc$ separately by using the Helmholtz decomposition that splits a vector field uniquely into a curl-free (longitudinal) and a divergence-free (transverse) component. The same approach was recently suggested in Ref.~\onlinecite{sharma2018source} in order to make a magnetic-field functional source-free. The decomposition for the internal-force terms is $\FHxc/n = -\nabla \varphi + \nabla \times \bm{\alpha}$ with the curl-free component $-\nabla\varphi$ and the divergence-free component $\nabla\times\bm{\alpha}$. If we choose to set $\vHxc = \varphi$, the vector potential has to fulfill the evolution equation
\begin{equation}\label{eq-Ax}
\begin{aligned}
n&\partial_t \Axc = \jj \times (\nabla \times \Axc) + (\Axc \otimes (\jj+n\AA) )\nablaleft  \\
&+((\jj+n\AA+n\Axc) \otimes \Axc)\nablaleft - n\nabla \times \bm{\alpha}[\Phi,\Psi].
\end{aligned}
\end{equation}
If we just neglect the correlation contributions from $\FHxc$ as in Eq.~\eqref{eq:vHxc-parts}, i.e., set $\FHxc[\Phi,\Psi] \approx \FF_W[\Phi]$, we get a corresponding version of Eq.~\eqref{eq-Ax}, where the exchange part of the vector potential $\bm{\alpha}_\mathrm{x}[\Phi]$ again only depends on the Slater-determinant wave function of the Kohn--Sham system,
\begin{align}
    &\FF_W[\Phi]/n = -\nabla \varphi_\mathrm{x}[\Phi] + \nabla \times \bm{\alpha}_\mathrm{x}[\Phi], \\[.5em]
    &\vHx = \varphi_\mathrm{x}[\Phi], \\
    &n\partial_t \Ax = \jj \times (\nabla \times \Ax) + (\Ax \otimes (\jj+n\AA) )\nablaleft \label{eq-AxEq}   \\
&+((\jj+n\AA+n\Ax) \otimes \Ax)\nablaleft - n\nabla \times \bm{\alpha}_\mathrm{x}[\Phi].\nonumber 
\end{align}
This yields the local-exchange approximation for both the scalar and the vector potential.

Yet the previous choice $\vHxc = \varphi$, by separating it from the vector potential as a unified field, amounts to a specific choice of gauge, a fact that must be noted carefully at this place. Complete gauge freedom was still retained for the choices of $(\vHxc,\Axc)$ up to Eq.~\eqref{eq-xc-potentials}. What is still left from the original gauge freedom is the possibility of adding a time-\emph{independent} gradient field to the vector potential.
Further, Eq.~\eqref{eq-AxEq} as an \emph{evolution}-equation approximation to $\Axc$ already includes memory effects. The xc vector potential captures the rotational contributions $\nabla\times\bm{\alpha}$ from the Hartree-exchange-correlation force term $\FHxc/n$ that are lost if only the exchange potential (from the longitudinal component $-\nabla\varphi$) is considered like in Eq.~\eqref{eq-Hxc-potential-2}.
That the determining relation for $\Axc$ takes the form of an evolution equation has tremendous advantages as any unbalanced part from the forces on the right hand side of Eq.~\eqref{eq-Ax} (i.e., a non-zero right hand side) can just be absorbed by the time-derivative $\partial_t \Axc$.

A further condition for $\Axc$ can be derived in terms of $\jpara$. Since one demands $\jj[\Phi]=\jj[\Psi]$ it holds by the definition of the physical current density \eqref{current_def} that
\begin{equation}\label{eq:remarkable-para}
  \jpara[\Phi]-\jpara[\Psi]= n\Axc.  
\end{equation}
Note further that since the initial states $(\Phi, \Psi)$ and the initial density fix $\vHxc$ (by fixing $\FHxc$) and $\Axc$ (by Eq.~\eqref{eq:remarkable-para} above) up to a gauge at the initial time, the evolution equation allows us to determine $\Axc$ (and approximations like $\Ax$) at the next time step uniquely by Eq.~\eqref{eq-xc-potentials}. 
By integration of Eq.~\eqref{eq-xc-potentials} over the whole space and application of Gauss' theorem like in Sec.~\ref{sec:eom-current} we arrive at the zero-force and zero-torque constraints (see Ref.~\onlinecite[Eq.~(22)]{tokatly2005quantum2}) that are always fulfilled by the exact pair $(\vHxc,\Axc)$ and serve as consistency checks for approximate potentials,
\begin{align}\label{eq-Ax-exact-constraint}
&\int n(\partial_t \Axc + \nabla \vHxc)  \d \rr =\int \jj \times (\nabla \times \Axc) \d \rr, \\
&\int n \rr\times (\partial_t \Axc + \nabla \vHxc)  \d \rr =\int \rr\times (\jj \times (\nabla \times \Axc)) \d \rr.
\end{align}

If one is interested in ground states, the determining relation for the xc potentials of Eq.~\eqref{eq-xc-potentials} holds just as well, with vanishing time derivative $\partial_t \Axc=0$. Although solving for $\Axc$ is now more involved than in the time-dependent case, we find an explicit orbital-dependent approximation for the pair $(\vHxc,\Axc)$. We also see that in the case of zero vector potentials, this approximation recovers the local-exchange approximation of Eq.~\eqref{eq:def-vHx}, while in the time-dependent case we recover the time-dependent local-exchange approximation of Eq.~\eqref{eq-td-Hx-potential} (if we solve only with the longitudinal forces).

Further, in TDCDFT, contrary to the stationary case, the availability of the continuity equation \eqref{continuity} makes it possible to get the density $n$ at every time from the fixed initial state and the physical current density $\jj$ by just integrating over time. This fits to the insight that gauge freedom of $\AA$ can be used in Eq.~\eqref{eq-force-balance} to completely remove $v$ (called the temporal gauge choice) as the complementary variable of the density.
Thus the quantity to match between the two systems is naturally the physical current density $\jj$. This would then amount to another gauge choice than the one we implicitly chose by the Helmholtz decomposition, setting $\vHxc=0$ and leaving the whole Hartree-exchange-correlation force in Eq.~\eqref{eq-Ax} instead of only its transverse part. One needs to keep in mind that since the continuity-equation connection between the one-particle density and the current density does not hold in the same way for the ground-state setting, there the two quantities cannot be possibly controlled by a vector field $\Axc$ alone and a scalar potential $\vHxc$ is needed.

\subsection{Paramagnetic CDFT xc potentials}
\label{Sec:paraCDFT_xc}

Next we consider the paramagnetic current density as a basic variable to get a defining equation for the xc potentials. This introduces a significant conceptual shift since for fixed density $n$ the currents $\jj$ and $\jpara$ are linked through subtraction of $n\AA$, but the vector potential can be different for the interacting and non-interacting system. So demanding matching physical currents $\jj$ for both systems as before creates a different rule for the xc (vector) potential than demanding identical paramagnetic currents $\jpara$.
That it is indeed possible to set up a non-interacting reference system that yields the same $(n,\jpara)$ as the interacting system relies on a weaker version of the Hohenberg--Kohn result that guarantees the existence of a (nondegenerate) ground-state wave-function $\Phi$ for given densities $(n,\jpara)$~\cite{Vignale-Rasolt-1987,Vignale-Rasolt-1988}, although the potentials $(v,\AA)$ cannot be determined~\cite{Capelle-Vignale-2002}. This also holds for degenerate ground-states, although the level of degeneracy cannot be determined by the densities~\cite{LaestadiusTellgren}.
In the time-dependent case an extension of the Runge--Gross result to the paramagnetic case is not available to the best of our knowledge. Again we merely assume the existence of such a map.
A detailed mathematical analysis of paramagnetic ground-state CDFT including a rigorous setup for constructing a Kohn--Sham iteration scheme that is in fact \emph{only} possible for the \emph{paramagnetic} current density was presented in Ref.~\onlinecite{MY-CDFTpaper2019}. From Eq.~\eqref{eq-force-balance-para} we know the EOM for both the interacting and the non-interacting reference system,
\begin{align}
\partial_t \jpara = &-n \nabla v + (\nabla \otimes \AA)(\jpara - n\AA) \nonumber\\
&+ (\jpara \otimes \AA)\nablaleft + \FF_T[\Psi] + \FF_W[\Psi], \label{eq:force-balance-para-int}
\\[.6em]
\partial_t \jpara = &-n \nabla v_s + (\nabla \otimes \AA_s)(\jpara - n\AA_s) \nonumber\\
& + (\jpara \otimes \AA_s )\nablaleft + \FF_T[\Phi]. \label{eq:force-balance-para-non-int}
\end{align}
We connect them in the same way as we did for Eq.~\eqref{eq-xc-potentials} previously to get
\begin{equation}\label{eq:xc-jpara}
\begin{aligned}
n&\nabla \vHxc = (\nabla \otimes \Axc)(\jpara - n\AA - n\Axc) \\
&- (\nabla \otimes \AA)n\Axc +(\jpara \otimes \Axc )\nablaleft 
- \FHxc[\Phi,\Psi].
\end{aligned}
\end{equation}
This gives us a new defining equation for $\vHxc$ and $\Axc$ for matched densities and paramagnetic currents. If we then use the same Helmholtz decomposition for $\FHxc/n$ as before, it provides $\vHxc = \varphi$ and
\begin{equation}\label{eq-Ax-para}
\begin{aligned}
    (\nabla \otimes \Axc) &(\jpara - n\AA - n\Axc) - (\nabla \otimes \AA) n\Axc \\
    &+ (\jpara \otimes \Axc) \nablaleft = n\nabla \times \bm{\alpha}[\Phi,\Psi].
\end{aligned}
\end{equation}
The corresponding local-exchange approximation becomes $\vHx = \varphi_\mathrm{x}[\Phi]$ and
\begin{equation}
\begin{aligned}
(\nabla \otimes \Ax) &(\jpara - n\AA - n\Ax) - (\nabla \otimes \AA) n\Ax  \\
&+(\jpara \otimes \Ax) \nablaleft = n\nabla \times \bm{\alpha}_\mathrm{x}[\Phi].
\end{aligned}
\end{equation}
Again, if we consider the limit of no external vector potentials we recover the local-exchange approximation of Eqs.~\eqref{eq:def-vHx} and~\eqref{eq-td-Hx-potential}. 

When matching densities and paramagnetic currents, a condition for $\Axc$ can be easily derived that was called a ``remarkable identity'' in Ref.~\onlinecite{vignale-rasolt-geldart}. Since $n[\Phi]=n[\Psi]$ also their time-derivatives agree and that means $\nabla \cdot \jj[\Phi] = \nabla \cdot \jj[\Psi]$ from the continuity equation \eqref{continuity}. If one assumes now the paramagnetic currents to be equal, this yields $\nabla \cdot (n\AA_s) = \nabla \cdot (n\AA)$ and consequently
\begin{equation}\label{eq:remarkable}
\nabla\cdot(n\Axc)=0.
\end{equation}
This gives a further condition for $\Axc$ that holds in the paramagnetic CDFT setting.
The new condition \eqref{eq:remarkable} can be seen as another form of gauge fixing that arrives together with a loss of gauge freedom for $\Axc$ that occurred already at the start by considering a quantity that is not gauge invariant, $\jpara$ in this case. This reduced freedom for a choice of $\Axc$ when matching paramagnetic currents might also be the reason why we do not arrive at an evolution equation like in the previous case, but at a rigid balance equation like Eq.~\eqref{eq:xc-jpara}. Note that Eq.~\eqref{eq:remarkable} would hold both for the ground-state and the time-dependent setting, once we match the paramagnetic current density and the one-particle density. In the ground-state setting it just additionally holds that $\nabla \cdot \jj[\Phi] = \nabla \cdot \jj[\Psi]=0$. One could try to fulfill Eq.~\eqref{eq:remarkable} in a time-dependent setting at every time instant by treating it as a gauge condition and by choosing a corresponding time-dependent gauge. But such a time-dependent gauge field $\Lambda(t)$ then influences the scalar potential by $-\partial_t \Lambda$ and there would be no hope to still simultaneously fulfill $\vHxc = \varphi$, where $\varphi$ is determined by the xc force. All these indications (together with the available proofs for density-potential maps) support the standard choices for the relevant current quantity, i.e., to employ the physical current density for a time-dependent setting and the paramagnetic current density for ground-state CDFT.

It was noted in Ref.~\onlinecite{Tellgren2012} that introducing a new scalar potential $u = v+\onehalf |\AA|^2$ is beneficial for giving a concave ground-state energy functional that fits into Lieb's convex-conjugate formulation~\cite{Lieb1983}. It also simplifies the expression for the xc potentials Eq.~\eqref{eq:xc-jpara} since the terms quadratic in $\AA, \AA_s$ from Eqs.~\eqref{eq:force-balance-para-int} and \eqref{eq:force-balance-para-non-int} drop out by using $\onehalf \nabla |\AA|^2 = (\AA\cdot\nabla)\AA+\AA\times(\nabla\times\AA) = (\nabla\otimes\AA)\AA$ from Eq.~\eqref{eq:vec-id-2}. With the new scalar xc potential $\uHxc = u_s-u$, Eq.~\eqref{eq:xc-jpara} becomes the much simpler
\begin{equation}\label{eq:u-xc-jpara}
n\nabla \uHxc = (\nabla \otimes \Axc)\jpara +(\jpara \otimes \Axc )\nablaleft 
- \FHxc[\Phi,\Psi].
\end{equation}

When considering a gauge change $\AA \to \AA + \nabla\Lambda$ (time-dependent or not) the paramagnetic current density transforms as $\jpara \to \jpara - n\nabla\Lambda$ as can be seen directly from the definition in Eq.~\eqref{current_def} and the fact that $n$ and $\jj$ are gauge independent. In focusing for a while on the quantity $\jpara/n$, we see that an arbitrary gradient field can be added to it, leaving the contribution from a rotational field as the physical, gauge-independent part. To delete the contributions from the fully gauge-dependent gradient field, one can apply the curl, which then leads to the (gauge-independent) vorticity,
\begin{equation}
\bm{\nu} = \nabla \times \frac{\jpara}{n}.
\end{equation}
The vorticity was considered as a basic variable for CDFT already in Ref.~\onlinecite{Vignale-Rasolt-1987}. When considering an xc potential that matches the paramagnetic current density and the density like before, then by definition also the vorticities of the systems agree.

\section{Local Density Approximation from Exchange Forces}
\label{sec:LDA}

We have found that all the different EOM-based local-exchange approximations reduce to the local-exchange approximations of Eqs.~\eqref{eq:def-vHx} and~\eqref{eq-td-Hx-potential} in the limit of no external vector potential, respectively. We do not know, however, how they correspond to well-established energy-based approximations. Only in the special case of two particles in a singlet configuration it is known that the EOM-based exchange approximation becomes equivalent to the usual energy-based exchange approximation \cite{ruggenthaler2009local}. To highlight the connection between both approximation schemes we will show in the following that the local-exchange approximation reduces to the well-known exchange-only LDA of ground-state DFT~\cite[Sec.~7.4]{parr} in the homogeneous case. The new form of derivation presented here even entails a direct connection to the related X$\alpha$ method that was introduced by Slater~\cite{slater1951} as a simplification to the Hartree--Fock method. The X$\alpha$ method introduces a parameter $\alpha$, sometimes interpreted as adjustable~\cite{slater1969comparison}, that takes a value between $2/3$ (reproducing exactly the usual LDA case) and $1$. In our derivation this parameter can be interpreted as the position-choice of a \emph{reference point} $\rr_\lambda$.

\subsection{\label{sec:level2.1}Single Slater Approximation}

The wave function of the ground state for a non-interacting system is commonly approximated by a Slater determinant of the $N$ lowest occupied spin orbitals $\phi_i(\xx)$,
\begin{equation}
\label{slater-det-wavefuncion}
\Phi = \frac{1}{\sqrt{N!}}\sum_{\sigma \epsilon S_{N}} \sign(\sigma)\phi_{\sigma(1)}(\xx_1)\ldots\phi_{\sigma(N)}(\xx_N),
\end{equation}
with $S_{N}$ the permutation group of $N$ elements. The 1-RDM of such a single Slater determinant wave-function is
\begin{equation}
\rho_{(1)}(\rr_1,\rr'_1) = \sum_{i=1}^{N} \sum_{s}\phi_{i}(\rr_1,s)\phi_{i}^*(\rr'_1,s)
\end{equation}
and the 2-RDM can in this case be expressed as a determinant of only the 1-RDM,
\begin{equation}\label{eq:slater-2RDM}
\rho_{(2)}(\rr_1,\rr_{2},\rr_1,\rr_{2}) = \frac{1}{2}\left( n(\rr_1)n(\rr_{2}) - \frac{1}{2}\left| \rho_{(1)}(\rr_1,\rr_{2})\right| ^2\right) .
\end{equation}
Replacing the latter in the interaction force term \eqref{eq-interaction-force} we get
\begin{widetext}
\begin{equation}
\begin{aligned}
\FF_W[\Phi] &= -\int (\nabla w(\rr_1-\rr_{2})) \left( n(\rr_1)n(\rr_{2}) - \frac{1}{2}\left| \rho_{(1)}(\rr_1,\rr_{2})\right| ^2\right) \d\rr_{2} \\
&= - n(\rr_1)\nabla\int w(\rr_1-\rr_{2})n(\rr_{2}) \d\rr_{2} +\frac{1}{2} \int (\nabla w(\rr_1-\rr_{2}))\left| \rho_{(1)}(\rr_1,\rr_{2})\right|^2 \d\rr_{2} \\[.7em]
&= \FH[\Phi] + \Fx[\Phi].
\end{aligned}
\label{eq-interaction-force-single-slater}
\end{equation}
\end{widetext}
Here the first term is simply the Hartree mean-field contribution to the force. We look into the second term to get the pure exchange effects and to be able to define an effective exchange potential functional $\vx$,
\begin{align}
    n\nabla \vH &= -\FH[\Phi] \label{eq:def-vH}\\
    &\Rightarrow \vH[n](\rr) = \int w(\rr_1-\rr_{2})n(\rr_{2}) \d\rr_{2}, \nonumber\\
    n\nabla \vx &= -\Fx[\Phi]. \label{eq:def-vx}
\end{align}

In order to move beyond exchange-only approximations, the expression for the 2-RDM would include more refined exchange-correlation approximations to the \emph{pair correlation function} that is simply $h(\rr_1,\rr_2) \approx -|\rho_{(1)}(\rr_1,\rr_{2})|^2 / (2 n(\rr_1)n(\rr_{2}))$ in Eq.~\eqref{eq:slater-2RDM}.

\subsection{\label{sec:level2.2}LDA derivation}

Next we consider the homogeneous limit, i.e., the uniform-electron-gas case of particles in a volume $V=L^3$ with periodic boundary conditions. This leads to orbitals of the form
\begin{equation}
\varphi_\kk(\rr) = \frac{1}{\sqrt{V}}\e^{\i \kk\cdot\rr}
\end{equation} 
with $\kk=(n_x,n_y,n_z)2\pi/L$, each one doubly occupied starting with the lowest absolute momentum $n_x,n_y,n_z=0,\pm1,\pm2,\ldots$ up to the Fermi-sphere surface $|p|=p_f=\hbar k_f$ with $k_f = (3\pi^2 n) ^\frac{1}{3}$ (see Ref.~\onlinecite[Eq.~(6.1.12)]{parr}; this is the ``closed shell'' assumption and requires an even number of electrons). The 1-RDM is then obtained as \cite[Sec.~6.1]{parr}
\begin{equation}
\begin{aligned}
\rho_{(1)}&(\rr_1,\rr_{2}) = \frac{1}{4\pi^3} \int_{|k|< k_f} \e^{\i \kk\cdot(\rr_1-\rr_2)} \d\kk \\
&= \frac{1}{2\pi^2}\int_{0}^{k_f} k^2 \d k \int_{0}^{\pi} \sin\theta \e^{\i k|\rr_1-\rr_2|\cos\theta} \d\theta \\
&= \frac{1}{2\pi^2}\int_{0}^{k_f} k^2 \left( \frac{2\sin(k|\rr_1-\rr_2|)}{k|\rr_1-\rr_2|}\right)  \d k \\
&= \frac{1}{\pi^2|\rr_1-\rr_2|}\int_{0}^{k_f} k \sin(k|\rr_1-\rr_2|) \d k \\
&= \frac{1}{\pi^2}\frac{\sin(k_{f}|\rr_1-\rr_2|) - k_{f}|\rr_1-\rr_2|\cos(k_{f}|\rr_1-\rr_2|)}{|\rr_1-\rr_2|^3}.
\label{LDA-1RDM}
\end{aligned}
\end{equation}
The integral over the sphere of radius $k_f=p_f/\hbar$ has been transformed to spherical coordinates $(r,\theta,\varphi)$, the $z$-axis pointing into the direction $\rr_1-\rr_2 $ and $\theta$ being the angle between the $\kk$-vector and $\rr_1-\rr_2 $. For an arbitrary density $n(\rr)$ the $k_f$ will depend on the position $\rr$,
\begin{equation}
k_f(\rr) = \left( 3\pi^2 n(\rr)\right) ^\frac{1}{3},
\label{fermi-radius}
\end{equation}
and the question arises at which point $\rr_\lambda$ we should take it in the integral for $\rho_{(1)}(\rr_1,\rr_{2})$ above, since it depends on two positions, $\rr_1$ and $\rr_2$.
We choose a weighted average of $\rr_1$ and $\rr_2$, that is, $\rr_\lambda=\rr_1+\lambda(\rr_2-\rr_1)=\rr+\lambda \rr'$.
Then we would like to express the 1-RDM in terms of $(\rr_{\lambda},|\rr'|)$.
Such a flexible choice of the average coordinate was suggested before in Ref.~\onlinecite{Koehl1996} for a density matrix expansion, but it only affects density-gradient terms there, while the parameter $\lambda$ will remain here also in the lowest order.
The substitutions $t=k_f(\rr_{\lambda})|\rr_1-\rr_2| = k_f(\rr_{\lambda})|\rr'|$ and \eqref{fermi-radius} put into Eq.~\eqref{LDA-1RDM} give
\begin{equation}
\rho_{(1)}(\rr_1,\rr_{2}) = 3n(\rr_\lambda)\frac{\sin t - t\cos t}{t^3}.
\end{equation}
We can use this expression to evaluate the exchange term in Eq.~\eqref{eq-interaction-force} for the usual Coulomb interaction potential $w(\rr_1-\rr_2) = |\rr_1-\rr_2|^{-1}$,
\begin{equation}
\begin{aligned}
\Fx &= \frac{1}{2}\int \left( \nabla\frac{1}{|\rr_1-\rr_2|}\right)  \left|\rho_{(1)}(\rr_1,\rr_{2})\right|^2 \d \rr_2\\
&= \frac{9}{2}\int \frac{\rr'}{|\rr'|^3} n(\rr_\lambda)^2 \left( \frac{\sin t -t\cos t}{t^3}\right)^2 \d \rr'.
\end{aligned}
\end{equation}
We transform this integral to spherical coordinates again, $\d \rr' = |\rr'|^2 \d |\rr'|\d \Omega = |\rr'|^2 \frac{\d t}{k_f} \d \Omega$ and $\d \Omega = \sin\vartheta \d \vartheta \d \varphi$, and have
\begin{equation}\label{eq:Fx-derivation}
\begin{aligned}
\Fx &= \frac{9}{2}\int \frac{\rr'}{|\rr'|^3}\frac{|\rr'|^2}{k_{f}} n(\rr_{\lambda})^2 \left( \frac{\sin t -t\cos t}{t^3}\right)^2 \d t \d\Omega \\
&=  \frac{9}{2}\int\frac{\rr'}{|\rr'|^2}\frac{n(\rr_{\lambda})^2}{(3\pi^2 n(\rr_{\lambda}))^{\frac{2}{3}}}\frac{\left(\sin t - t\cos t\right)^2}{t^5}\d t \d\Omega \\
&=  \frac{9}{2(3\pi^2)^{\frac{2}{3}}} \int\frac{\rr'}{|\rr'|^2} n(\rr_{\lambda})^{\frac{4}{3}}\frac{\left(\sin t - t\cos t\right)^2}{t^5}\d t \d\Omega
\end{aligned}
\end{equation}
Since the LDA exchange energy depends on the local density, which is assumed to be slowly varying, it seems natural to relate the exchange force to the local gradient of the density term.
We thus make a Taylor expansion of $n(\rr_{\lambda})^{\frac{4}{3}}$ at $\rr_1$,
\begin{equation}\label{eq:taylor-expansion}
n(\rr_{\lambda})^{\frac{4}{3}} = n(\rr_1)^{\frac{4}{3}} + \frac{4}{3}\lambda \rr' (\nabla n(\rr_1))n(\rr_1)^{\frac{1}{3}} + \ldots.
\end{equation}
Then all even terms of the Taylor expansion vanish in Eq.~\eqref{eq:Fx-derivation} because of symmetry in the integral over $\rr'$. We further limit ourselves to the first-order approximation in Eq.~\eqref{eq:taylor-expansion} only.
\begin{widetext}
\begin{align}
&\Fx(\rr_1) = n(\rr_1) \nabla\vx[n](\rr_1) \approx \frac{9}{2(3\pi^2)^{\frac{2}{3}}} \int\frac{\rr'}{|\rr'|^2} \frac{4}{3}\lambda \rr' (\nabla n(\rr_1))n(\rr_1)^{\frac{1}{3}} \frac{\left(\sin t - t\cos t\right)^2}{t^5}\d t \d\Omega \nonumber\\
&= \frac{9}{2} \frac{4\lambda n(\rr_1)^{\frac{1}{3}}}{3(3\pi^2)^{\frac{2}{3}}}\underbrace{ \int \frac{\rr'}{|\rr'|} \left( \frac{\rr'}{|\rr'|}\cdot\nabla n(\rr_1)\right)\d \Omega}_{\frac{4\pi}{3}\nabla n(\rr_1)}  \underbrace{\int\frac{\left(\sin t -t\cos t\right)^2}{t^5}  \d t}_{\frac{1}{4}} = \frac{2}{3} \lambda \left(\frac{3}{\pi}\right)^\frac{1}{3} n(\rr_1)^\frac{1}{3} \nabla n(\rr_1) =  2 \lambda \left(\frac{3}{\pi}\right)^\frac{1}{3} n(\rr_1) \nabla n(\rr_1)^\frac{1}{3}
\label{lda-force-term}
\end{align}
\end{widetext}
Here we have used that the vector projection of an arbitrary vector $\mathbf{a}$ yields 
\begin{equation}
\int\frac{\rr'}{|\rr'|} \frac{\rr' \cdot \mathbf{a}}{|\rr'|} \d\Omega = \frac{4\pi}{3}\mathbf{a}.
\end{equation}
We can now just read off the local-density approximation from Eq.~\eqref{lda-force-term} as the first order term of the local-exchange potential and compare it to the usual LDA expression $\vx^{\mathrm{LDA}}$ (see Ref.~\onlinecite[Eqs.~(7.4.2) and (7.4.5)]{parr}),
\begin{align}
\vx^{\lambda\mathrm{LDA}}[n](\rr) &= -2 \lambda \left(\frac{3}{\pi}\right)^\frac{1}{3} n(\rr)^\frac{1}{3} = 2\lambda \vx^{\mathrm{LDA}}[n](\rr).
\end{align}
If one chooses $\lambda = 1/2$, which means taking the density at the middle point between $\rr_1$ and $\rr_2$, exactly like in the usual energy-based derivation (see Ref.~\onlinecite[Eq.~(6.1.13)]{parr}), this yields the usual LDA exchange-only potential. 

However, there exist variants of LDA for the exchange potential, one of which is called the \textit{Slater exchange} $\vxalpha[n]$ \cite{slater1951}. It was proposed as a simplification of the Hartree--Fock method, where the uniform-electron-gas model was used as well, resulting in the expression,
\begin{equation}
\label{X-alpha-eq}
\vxalpha[n] = \frac{3}{2}\alpha \vx^{\mathrm{LDA}}[n]
\end{equation}
where $\alpha$ was originally set to 1. The competing values, $\alpha=1$ and $\alpha=2/3$ for the usual LDA, come from applying the uniform-electron-gas approximation at different stages of the derivation. This ambiguity has led to an attitude where $\alpha$ was taken as an adjustable parameter that can be fitted to empirical evidence \cite{slater1969comparison}.
We note that in our derivation such a variable parameter, namely $\lambda=3\alpha/4$, occurs as well, but with a clear conceptual meaning in the course of the derivation, i.e., the location around which the Taylor expansion \eqref{eq:taylor-expansion} is taken.

\section{Conclusion}
\label{sec:conclusions}

In this work we have investigated an approach towards the xc potentials of various density-functional theories based on specific current density EOMs instead of energy or action expressions. In this way we establish exact, pointwise determining relations for the xc potential that allow for straightforward orbital-dependent approximations. This approach has the advantage over energy- and action-based approximation strategies in that no OEP procedure needs to be employed and subtle mathematical and conceptual issues can be avoided. Furthermore, we highlighted many details about the connections between the presented local-exchange approximations for (TD)DFT and (TD)CDFT, such as differences between ground-state DFT and TDDFT due to transverse local forces and gauge-fixing that stems from the separation of scalar and vector xc potential. Finally, we showed that in the limit of a homogeneous system the local-exchange approximations reduce to the usual exchange LDA and provided an intuitive connection to the X$\alpha$ method.    

By avoiding the usual route over an energy functional in the static case, we lose the direct access to total energies which are of prime interest in most quantum chemistry applications. Moreover, it is not clear apriori if for an xc potential that is obtained from an EOM-based approximation such a corresponding parent energy functional even exists. Nevertheless, total energies can be obtained aposteriori from the converged Kohn--Sham orbitals using for example the same ansatz for the 2-RDM or the interacting wavefunction that was used to approximate $\FF_T[\Psi]$ and $\FF_W[\Psi]$ of Eq.~\eqref{eq-Hxc-potential}.

To improve upon the local-exchange approximations, those internal force terms $\FF_T[\Psi]$ and $\FF_W[\Psi]$ approximated in terms of the Kohn--Sham wave function $\Phi$ need to be studied in more detail for the different cases. While they already have been quite extensively discussed in terms of densities and currents~\cite{Vignale-Kohn1996,tokatly2005quantum2}, their orbital-dependent expressions are much less studied. The explicit approximation of the involved reduced quantities, the 1-RDM and the diagonal of the 2-RDM, similar to considerations about the xc hole~\cite{PribramJones2015,Giesbertz2013}, is a possible route. Working with auxiliary equations that determine these objects similar as in Ref.~\onlinecite{lacombe2019density} seems an interesting way to provide approximations for the correlation parts of the xc potentials too. The latter approach is closely connected to the idea of working with different density quantities. Although the full 1-RDM of an interacting system is in general not reproducible by a Kohn--Sham wave function $\Phi$, other more complex reduced quantities than the density and the currents might still be. For instance, on a lattice the link-current together with the kinetic-energy density constitute a possible alternative to current- or density-based time-dependent functional theories~\cite{CDFT_Lattice}. Another example would be the vorticity or a tensorial unification of vorticity and kinetic-energy density that was recently proposed \cite{Sen2018} as fundamental functional variables. In such cases also the external fields need to be adopted and potentially take a more complex, tensorial form, but the EOM-based strategy to devise approximations presented here can be straightforwardly adopted. This will be the subject of future work. Besides, the strategy can be easily extended to other physical settings, such as matter-photon density-functional theories relating to quantum-electrodynamical systems~\cite{QEDFT,Ruggenthaler2011,Tokatly2013,Ruggenthaler2014}.  

\begin{acknowledgments}
We express our gratitude for helpful comments from Andre Laestadius on an earlier draft and for insightful discussions with Florian Eich.
This project has received funding from the European Research Council (ERC) under the European Union’s Horizon 2020 research and innovation programme (Grant agreement No.~694097), as well as from the Deutsche Forschungsgemeinschaft (DFG) Sonderforschungsbereich 925 ``Light-induced dynamics and control of correlated quantum systems'' (project No.~A4) and the DFG Cluster of Excellence ``Advanced Imaging of Matter'' (EXC 2056, project ID 390715994).
M.P.\ acknowledges support by the Erwin Schr\"odinger Fellowship J 4107-N27 of the FWF (Austrian Science Fund).
\end{acknowledgments}

\end{document}